\documentclass[aps,floats,prl,
showpacs, twocolumn ]{revtex4}

\usepackage{amsfonts,amsmath,mathrsfs} \usepackage{bm} \usepackage{dcolumn}
\usepackage{graphicx}
 \usepackage{latexsym}

\begin{document}

\title{Unusual Brownian motion of photons in open absorbing media}

\author{Li-Yi Zhao$^1$, Chu-Shun Tian$^1$, Zhao-Qing Zhang$^2$ and Xiang-Dong Zhang$^3$}

\affiliation{$^1$Institute for Advanced Study, Tsinghua University, Beijing 100084, China\\
$^{2}$Department of Physics and William Mong Institute of Nano Science and Technology, Hong Kong University of Science and
Technology, Clear Water Bay, Kowloon, Hong Kong\\
$^3$School of Physics, Beijing Institute of Technology, Beijing 100081, China}

\begin{abstract}
{\rm Very recent experiments have discovered that
localized light in strongly absorbing media displays intriguing diffusive phenomena.
Here we develop a first-principles theory of light propagation in
open media with arbitrary absorption strength and sample length. We show analytically
that photons in localized open absorbing media exhibit
unusual Brownian motion. Specifically, wave transport follows the
diffusion equation with the diffusion coefficient exhibiting spatial
resolution. Most strikingly, despite that the system is controlled by two parameters -- the ratio of
the localization (absorption) length to the sample length --
the spatially resolved diffusion coefficient displays novel
single parameter scaling: it depends on the space
via the returning probability.
Our analytic predictions for this diffusion coefficient
are confirmed by numerical simulations. In the strong absorption
limit they agree well with the experimental results.
}
\end{abstract}
\maketitle

Many complex systems exhibit fundamentally different physics
in various scales. A canonical example,
thoroughly studied by Einstein in one of his revolutionary papers in 1905 \cite{Einstein1905},
is the motion of pollens suspended in water (the Brownian motion).
In short time scales, pollens undergo bombardment by water molecules
and exhibits random motion. The resulting complexity of microscopic dynamics notwithstanding,
in large time scales emerges substantially simpler physics:
the evolution of the pollen concentration profile fully characterizes
the dynamics and is governed by a diffusion equation.
Such {\it emergent} diffusive phenomenon
is universal, and microscopic dynamics
only affect the diffusivity namely the value of the diffusion coefficient.

The classical electromagnetic wave in infinite random media is
the very system where universal macrosocpic diffusive phenomena emerge from
complicated microscopic dynamics. In short (time) scales,
the system's physics is dominated by light scattering off random dielectric field
described by the Maxwell equation. In scales larger than
the elastic mean free time, as noticed by astrophysicists long time ago \cite{Chandrasekhar},
light propagation may be viewed as
random motion of Brownian particles, the photons.
Yet, light displays the particle-wave duality, and
the wave nature has far-reaching consequences.
Specifically, wave interference
suppresses the diffusivity \cite{John84}.
In the presence of strong dielectric fluctuations,
the (bare) diffusion coefficient is strongly renormalized and
vanishes in a much larger time scale.
This is a hallmark of Anderson localization of light
(see Ref.~\cite{Lagendijk09} for a review), and the corresponding time scale
is called the localization time.

In realistic experiments or photonic devices the medium is open:
photons escape the medium from the interfaces. Therefore, diffusivity
of waves cannot be fully suppressed even disorders are strong.
A fundamental question thereby is: do localized waves in open media
exhibit certain macroscopic diffusive phenomena?
This is a long-standing issue (see Ref.~\cite{Tian13} for a recent review).
Conceptually, a prominent difficulty arises from the following.
On the one hand long-time transport of localized waves in open media
is dominated by rare long-lived, high transmission (resonant) states
\cite{Azbel83,Freilikher04}, while
on the other hand macroscopic diffusion of waves -- if exists -- characterizes
statistical behavior of wave propagation: these are two seemingly contradictory facets.
In fact, experiments and numerical simulations \cite{Zhang09,Tian10,Skipetrov10} have
shown that due to the failure of capturing resonant states
the prevailing macroscopic diffusion model
developed in Ref.~\cite{Lagendijk00} cannot describe
transport of strongly localized waves in open media.
This had inspired the opinion that the resonant state
and macroscopic diffusion of photons are two incompatible concepts.

Contrary to this common belief, the results achieved recently \cite{Tian10,Tian13} show that
the rare resonant states
do not wash out macroscopic diffusion of photons.
Rather, they are reconciled in the way that the spatial resolution of the
diffusion coefficient exhibits a novel scaling. Specifically, the diffusion coefficient depends on the space
via the returning probability. (One may show -- both analytically and numerically \cite{Tian10} --
that this novel scaling is completely
missed in the prevailing model \cite{Lagendijk00} due to
the unjustified self-consistent treatments employed there.)
This novel scaling was analytically found by using a first-principles theory \cite{Tian10,Tian08}
and has been fully confirmed by numerical simulations \cite{Tian10}.


So far we have focused on passive media. Yet,
absorption inevitably exists in realistic environments \cite{Genack11}.
In open media the strong interplay between absorption and localization
has been experimentally shown to cause surprising phenomena \cite{Genack06},
and may even find practical applications in
optical devices \cite{Freilikher03}.
Very recently, both realistic and numerical experiments have observed that
localized waves exhibit
unusual macroscopic diffusion in open absorbing media \cite{Yamilov13a,Yamilov13}.
However, these experiments (both realistic and numerical)
focus on short samples and strong absorption.
For such systems resonant states are unimportant.
In the opposite limit, where samples are sufficiently long and
absorption is weak, resonant states dominate transport of localized waves.
As these states are very sensitive to absorption due to long lifetimes \cite{Freilikher95},
a very fundamental issue arises:
{\it how do absorption interplay with macroscopic diffusion of localized waves
in open media?} The purpose of this Letter
is to present the first systematic first-principles study
of this issue for arbitrary sample length and absorption strength.

{\it Main results and experimental relevance.} ---
Specifically, we consider propagation of light in quasi-one-dimensional
open (uniformly) absorbing media of length $L$. This system is controlled by two
dimensionless parameters, $\xi/L$ and $\xi_a/L$, where $\xi$ ($\xi_a$)
is the localization (absorption) length. We find that -- even in
localized samples ($\xi/L\ll 1$) --
photons exhibit macroscopic diffusion irrespective of the
absorption rate $\gamma$. More precisely, in the presence of
a steady monochromatic light source located at $x'$ the wave intensity (namely energy density) profile,
${\cal Y}(x,x')$, obeys
\begin{equation}\label{eq:1}
    \left\{\gamma-\partial_x D(x)\partial_x\right\}{\cal Y}(x,x')=\delta(x-x').
\end{equation}
It resembles the normal diffusion equation,
but, the diffusion coefficient exhibits
a number of anomalies.
First of all, it has a spatial resolution, $D(x)$,
decreasing monotonically from the air-medium interface to the sample
mid-point and varying over several orders.
Most strikingly, despite that the system is controlled by two parameters,
$D(x)$ exhibits a novel single parameter scaling,
\begin{equation}\label{eq:22}
    D(x)/D(0)=D_\infty(\lambda(x))
\end{equation}
similar to the passive medium case. Here, the scaling factor $\lambda(x)$ is proportional to
the (static) returning probability which depends on
$x/L$ as well as $\xi/L$ and $\xi_a/L$, and
the scaling function $D_\infty(\lambda)$ is the same as that of
passive media. (The explicit form of both $\lambda(x)$ and $D_\infty(\lambda)$ will
be given below.) Bearing these anomalies and the analogy to the normal
diffusion equation, Eq.~(\ref{eq:1}) describes
`unusual Brownian motion' of photons.
Furthermore, from Eq.~(\ref{eq:22}) we are able to predict analytically
the spatial resolution of the diffusion coefficient
which is confirmed by numerical experiments.
We stress that our analytic results
are very general: they are valid for arbitrary sample length and absorption strength
and, additionally, for both time-reversal (orthogonal symmetry)
and broken time-reversal (unitary symmetry) systems.

In the case of strong absorption, our analytic results are simplified
and fully agree with the recent experimental findings shown in Refs.~\cite{Yamilov13a,Yamilov13}.
Specifically, if
absorption is strong such that $\xi_a/L\ll 1$, we find that $D(x)$ exhibits
a plateau in the regime ${\rm min}(x,L-x)\gtrsim \xi_a$.
Moreover, the plateau value is determined by single parameter, $\xi_a/\xi$,
\begin{equation}\label{eq:4}
    D(L/2)/D(0)=D_\infty(\xi_a/2\xi)
\end{equation}
(cf. the first line of Table \ref{tab:1} and \ref{tab:2}).
We stress that the plateau as well as Eq.~(\ref{eq:4})
exists only if $\xi_a/L\ll 1$.

\begin{figure}[h]
  \centering
\includegraphics[width=8.0cm]{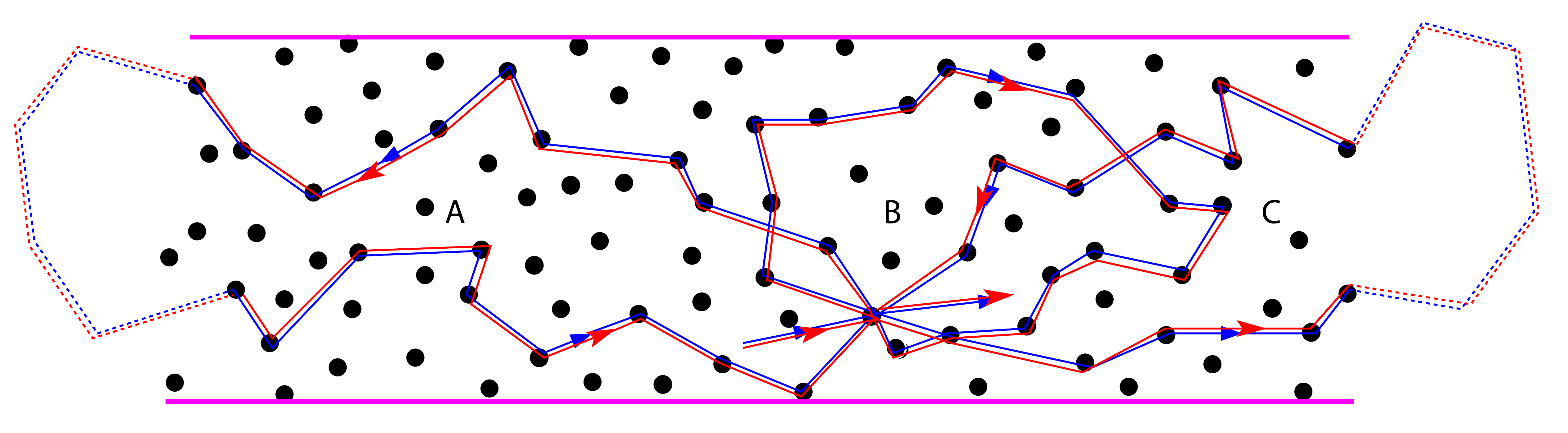}
 \caption{Examples of wave interference between two optical paths (red and blue)
 leading to the novel scaling behavior.
 The paths trace the three loops with different orders
 (B$\rightarrow$A$\rightarrow$C for the red and
 C$\rightarrow$B$\rightarrow$A for the blue).
 Notice that these loops may not be complete (e.g., dashed parts of A and C) due to
 photon leakage.}
  \label{fig:local}
\end{figure}

{\it Origin of unusual Brownian motion.} --- We begin with a physical
explanation of the main results. Since closer to the interface
more easily escape photons from the medium (cf. dashed lines of Fig.~\ref{fig:local}),
the returning probability, ${\cal Y}_0(x,x)$, is inhomogeneous in
space (i.e., depending on $x$ as well as
the parameters $\xi/L$ and $\xi_a/L$). As such,
wave interference effects are inhomogeneous also.
Near the interface (deep inside the sample)
they are weak (strong). For diffusive samples ($\xi/L\gg 1$ where
$\xi \sim \pi \nu D_0$ \cite{Efetov97}
with $\nu$ the photon density of states), ${\cal Y}_0(x,x)$ gives rise to
an inhomogeneous (one-loop) weak localization correction
$\sim {\cal O}({\cal Y}_0(x,x)/(\pi\nu))
\sim {\cal O}(\lambda (x))$.

For localized samples ($\xi/L\ll 1$), as waves penetrate deeply into the sample
photons easily return to its departure point.
This causes more complicated wave interference
essential to strong localization physics.
For example, two optical paths
may take the same $n$($=3$ in Fig.~\ref{fig:local}) loops
as their routes and trace them with different orders.
Therefore, they
constructively interfere with
each other. This leads to a weak localization correction
$\sim {\cal O}(\lambda^n (x))$. Therefore,
the spatial resolution $D(x)$ depends on
$x/L$ (as well as $\xi/L$ and $\xi_a/L$) via the
factor $\lambda(x)$, justifying Eq.~(\ref{eq:22}).

Importantly, the optical paths in Fig.~\ref{fig:local}
may propagate in the same direction -- clockwise or counterclockwise --
during tracing each loop, suggesting that interference picture beyond what leads to
usual one-loop weak localization does not necessarily require
time-reversal symmetry. This is crucial to localization physics of systems with
unitary symmetry (which is completely beyond the reach of
the self-consistent theory \cite{Lagendijk00}.) Therefore,
the novel single parameter scaling is an intrinsic wave phenomenon
unrelated to time-reversal symmetry.

{\it First-principles theory.} ---
The derivations are largely parallel to those of passive media \cite{Tian10,Tian08}.
Therefore, we shall outline the key steps below
with an emphasis on the key differences, and refer the readers to
Ref.~\cite{Tian13} for technical details.
We first introduce the microscopic formalism valid for arbitrary dimensions.
Consider the point-like source
located at ${\bf r}'$ inside the medium
with the spectral decomposition $J_{\omega}({\bf r}')$
($\omega$ being the wave angular frequency).
The time-integrated wave intensity -- upon
disorder averaging -- is given by \cite{Chernyak92}
$I({\bf r})=\int\frac{d\omega}{2\pi}
{\cal Y}({\bf r},{\bf r}'
)|J_{\omega
}({\bf r}')|^2
$. Here the (static) spatial correlation function
${\cal Y} ({\bf r},{\bf r}'
)
=\langle G^A_{\omega^2
} ({\bf r},{\bf
 r}') \, G^R_{\omega^2
 } ({\bf r}',{\bf r})\rangle$,
with $\langle\cdots\rangle$ standing for the disorder average
and the advanced (retarded) Green function $G^A$ ($G^R$) defined as
\begin{equation}
[\nabla^2 +\omega^2 (1+\epsilon ({\bf r})\pm i\epsilon'')]G^{R,A}_{\omega^2}({\bf r},{\bf
 r}')=\delta({\bf r} -{\bf r}'),
\label{Greenfunction}
\end{equation}
where the dielectric fluctuation, $\epsilon ({\bf r})$, is gaussian,
$\epsilon''>0$ causes uniform absorption, and the light velocity in
the air is set to unity.

Then, it is a standard procedure to
cast the above spatial correlation function
in terms of the functional integral over a supermatrix field
$Q({\bf r})$,
\begin{eqnarray}
{\cal Y} ({\bf r},{\bf r}') =\left(\frac{\pi \nu}{8\omega}\right)^2
\int D[Q]{\rm str} (A_+ Q({\bf r})A_-Q({\bf r}')) e^{-F[Q]}.
\label{DC}
\end{eqnarray}
Here $A_\pm$ are some constant supermatrices
and `str' the supertrace \cite{Tian13,Efetov97}. The action,
\begin{equation}
F[Q] = \frac{\pi\nu}{8} \, \int d{\bf r} {\rm str}\, \{D_0(\nabla Q)^2
-2\gamma\Lambda Q\},
\label{action}
\end{equation}
differs from the passive one \cite{Tian13} in the second term
accounting for the absorption.
Notice that the (bare) diffusion coefficient
$D_0\equiv D(0)$ and for $\epsilon''\ll 1$ the absorption rate
$\gamma=\omega\epsilon''$ \cite{footnote1}.
The supermatrix $Q=T^{-1}\Lambda T$ where
$\Lambda$ is the so-called metallic saddle point and $T$
takes the value from the coset space of $UOSP(2,2|4)/UOSP(2|2)\otimes UOSP(2|2)$
for orthogonal symmetry and of $U(1,1|2)/U(1|1)\otimes U(1|1)$
for unitary symmetry. It is
very important that if the air-medium interface
is transparent, the supermatrix field is fixed to be $\Lambda$ at the interface.
This reflects that on the interface there is no photon accumulation \cite{Tian10,Tian13,Tian08}.

In quasi one dimension the microscopic formalism is
simplified. Specifically, the supermatrix $Q$ field
is homogeneous in the transverse plane.
$F[Q]$ reduces to
$\frac{\pi\nu}{8} \, \int_0^L dx {\rm str}\, \{D_0(\partial_x Q)^2
-2\gamma\Lambda Q\}$,
($\nu$ includes a factor of the cross section area.)
and the boundary condition to
\begin{equation}\label{eq:21}
    Q(x=0)=Q(x=L)=\Lambda.
\end{equation}
As a result, ${\cal Y}$ depend only on
the longitudinal coordinate, $x$.
Then, we follow the procedures of Refs.~\cite{Tian13,Tian08} to calculate ${\cal Y}(x,x')$
explicitly. We find that ${\cal Y}(x,x')$
satisfies Eq.~(\ref{eq:1}) with the boundary condition:
${\cal Y}(x=0,x')={\cal Y}(x=L,x')=0$. Furthermore,
we find that the diffusion coefficient has a spatial resolution $D(x)$,
and the latter is a functional of the factor $\lambda(x)={\cal Y}_0(x,x'=x)/(\pi\nu)$.
This justifies Eq.~(\ref{eq:22}).
The correlator ${\cal Y}_0(x,x')$
solves the normal diffusion equation:
$(\gamma-D_0\partial_x^2){\cal Y}_0(x,x')=\delta(x-x')$
implemented by the boundary condition ${\cal Y}_0(x=0,x')
={\cal Y}_0(x=L,x')=0$. Solving this equation we obtain
\begin{eqnarray}\label{eq:6}
\lambda(x) = \frac{\xi_a}{2\xi}\frac{\cosh(L/\xi_a)
-\cosh((L-2x)/\xi_a)}{\sinh(L/\xi_a)},
\end{eqnarray}
where the diffusive absorption length $\xi_a\equiv
\sqrt{D_0/\gamma}$.

For $\lambda\ll 1$ we find that the perturbative expansion of the scaling function
$D_\infty(\lambda)$ is identical to that at $\gamma=0$.
This implies that $D_\infty(\lambda)$ is
the same as the one at $\gamma=0$.
On the other hand, at $\gamma=0$ the scaling function has been found
analytically and fully confirmed by numerical experiments \cite{Tian10},
which has the following asymptotic form:
\begin{eqnarray}
\label{eq:19}
  D_\infty(\lambda) \sim \bigg\{\begin{array}{c}
                            1+c_1 \lambda +c_2 \lambda^2 +\cdots,\quad \lambda \ll 1,  \\
                            e^{-\lambda},\quad \lambda \gtrsim 1.
                          \end{array}
\end{eqnarray}
Notice that the coefficients $c_i$ depend on the system's symmetry.
In particular, for orthogonal symmetry $c_1 < 0$
while for unitary symmetry $c_1=0,\,c_2 < 0$.

Eqs.~(\ref{eq:1}), (\ref{eq:22}), (\ref{eq:6}) and (\ref{eq:19})
constitute a complete description of unusual Brownian motion
of photons in open absorbing media.
In essence, it is a macroscopic phenomenon emerging from
the microscopic Helmholtz equation (\ref{Greenfunction})
in large time scales. The present macroscopic theory
differs from the previous one for passive media \cite{Tian10,Tian13}
in the absorption term and that the spatial resolution
$D(x)$ depends also on a new parameter $\xi_a/L$.
Below we provide numerical
evidence of this intriguing phenomenon.

\begin{figure}[h]
  \centering
\includegraphics[width=8.0cm]{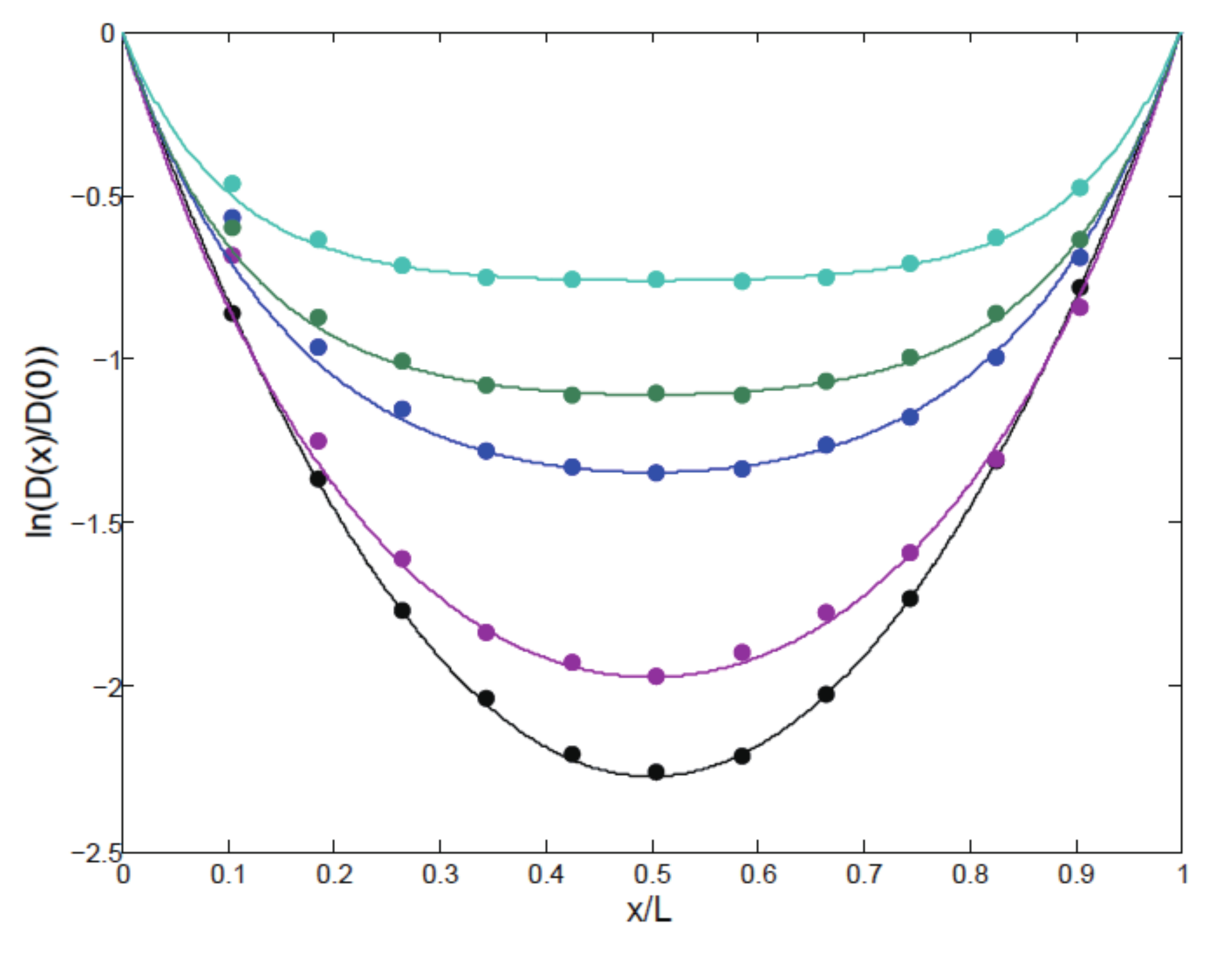}
 \caption{In localized open absorbing media the spatial resolution of
 the diffusion coefficient exhibits novel single parameter scaling.
 Analytic predictions for $D(x)$ (solid lines) are in good agreement with
 the simulation results (circles). From bottom to top
 the absorption strength increases (see the text for the value of $\epsilon''$
 as well as other parameters).}
  \label{fig:Dx}
\end{figure}

{\it Evidence from numerical experiments.} ---
We perform numerical simulations
of the (microscopic) Helmholtz equation. To this end we prepare
a randomly layered medium where the analytic results remain valid.
The relative permittivity in each layer is a random number
with a real part uniformly distributed in the interval $[0.3, 1.7]$.  The imaginary part,
$\epsilon''$, is assumed to be a constant. The random medium is embedded in the air background
so that there exist no internal reflections when $\epsilon''$ vanishes.
We launch a plane wave of frequency $\omega=1.65 a^{-1}$
into the system, where the sample length $L=200a$
with $a$ the layer thickness, and use the standard transfer matrix
method to calculate the wave intensity
profile, $I_\beta(x)$, for each dielectric disorder configuration $\beta$.
For a given value of $\epsilon''$, we calculate
the ensemble-averaged wave intensity
profile, $I(x)\equiv \langle I_\beta(x)\rangle$,
of $5\,000\,000$ disorder configurations.
Then, by presuming Eq.~(\ref{eq:1}), which gives
\begin{equation}\label{eq:3}
    D(x)=\frac{-\gamma \int_x^L I(x')dx'+D(0)\partial_x|_{x=L} I(x)}{\partial_x I(x)},
\end{equation}
we obtain numerical results of $D(x)$ from the measurements of $I(x)$.

The results of simulations are shown in Fig.~\ref{fig:Dx} for
$\epsilon''=0,1\times 10^{-4},\,5\times 10^{-4},\, 1\times 10^{-3}$, and
 $2\times 10^{-3}$ (from bottom to top).
First of all, in the passive limit ($\epsilon''=0$, the bottom curve) we recover
the result of Ref.~\cite{Tian10} for deeply localized
(Indeed, in this limit $L/\xi = 200a/22a\approx 9$ is large.) samples.
The effects of absorption become
significant when $\xi_a/L \sim {\cal O}(1)$, i.e.,
$\epsilon'' \approx \xi/(L^2\omega) \approx 3\times 10^{-4}$.
Having this estimation in mind, we may adjust the value of
$\epsilon''$ to systematically explore the effects of absorption.
The results are compared with the analytic predictions
obtained from the novel single parameter
scaling theory namely Eqs.~(\ref{eq:22}), (\ref{eq:6}) and (\ref{eq:19}).
As shown in Fig.~\ref{fig:Dx}, they are in good agreement.
Now we turn to analytic discussions of the behavior of $D(x)$ based on
the developed first-principles theory.

{\it Behavior of $D(x)$ in semi-infinite samples.} --- In this case $L\rightarrow \infty$, the scaling
factor (\ref{eq:6}) is simplified to
$\lambda(x)=\frac{\xi_a}{2\xi}(1-e^{-2x/\xi_a})$.
If the absorption is weak such that $\xi/\xi_a \ll 1$,
then $\lambda \ll 1$ for $x\ll \xi$ and $\lambda \gtrsim 1$ for $x\gtrsim \xi$.
Applying the first and second line of Eq.~(\ref{eq:19}), respectively,
we find distinct behavior of $D(x)/D_0$
in various regimes, which is summarized in
Table~\ref{tab:1}.
The first line indicates that
$D(x)$ reaches an exponentially small residual value at $x\gtrsim \xi_a/2$.
This reflects that deep inside the medium the ordinary
Brownian motion of photons is restored, albeit with
dramatically smaller diffusion coefficient. Qualitatively,
the plateau arises from that due to absorption
photons could not propagate for a distance larger than $\xi_a$ and
as such, photons deep inside the medium could not `see' the air-medium interface.
Indeed, similar results
were achieved by using the scaling theory of light
localization in infinite absorbing media long time ago \cite{John84}.
The second and third lines indicate an intermediate scale,
$\sqrt{\xi_a \xi}$, between $\xi$ and $\xi$.
For $x$ smaller than this scale the local diffusion coefficient
behaves essentially the same as that of (semi-infinite) passive media
(see the third and fourth lines),
while the deviation starts at $x\sim \sqrt{\xi_a \xi}$.
A physical explanation for this deviation may be as follows.
Consider light incident from the interface. It
effectively penetrates into the medium with a depth $\xi_a$. As such,
the Lyapunov exponent (inverse localization length), ${\bar \gamma}$,
fluctuates, following a distribution
$\sim e^{-\frac{\xi_a\xi}{4}({\bar \gamma}-\xi^{-1})^2}$ \cite{Anderson80}.
Averaging $e^{-{\bar \gamma} x}$ with respect to this distribution
we recover the result given in the second line.


If absorption is strong such that $\xi/\xi_a \gg 1$,
then $\lambda\ll 1$ irrespective of $x$.
By using the first line of Eq.~(\ref{eq:19}) we find the results
summarized in Table~\ref{tab:2}.

\begin{table}
  \centering
  \caption{Behavior of $D(x)$ at $\xi \ll \xi_a \ll L$}
  \label{tab:1}
\begin{tabular}{|l|l|}
    \hline
    regime & \qquad $D(x)/D(0)$   \\
\hline\hline $x\gtrsim \xi_a/2$ & \qquad $e^{-\xi_a/(2\xi)}$ \\
\hline $\sqrt{\xi_a \xi}\ll x\ll \xi_a/2$ & \qquad $e^{x^2/(\xi_a\xi)}e^{-x/\xi}$\\
\hline $\xi\lesssim x\ll \sqrt{\xi_a \xi}$ & \qquad $e^{-x/\xi}$ \\
\hline $x\ll \xi$ & \qquad $1+c_1 (x/\xi)+c_2 (x/\xi)^2+\cdots$ \\
    \hline
  \end{tabular}
\end{table}

{\it Behavior of $D(x)$ in finite localized samples.} --- These samples
($\xi/L\ll 1$) have an essential difference from
semi-infinite samples in the
existence of resonant states most of which reside near the sample
center \cite{Azbel83,Freilikher04}. As shown in Ref.~\cite{Tian10},
they play decisive roles in establishing
the highly unconventional macroscopic diffusion of localized waves.
Due to long lifetime the resonant state is very sensitive to absorption.
Below we shall discuss separately the cases of
weak ($\xi_a/L \gg 1$) and strong ($\xi_a/L \ll 1$) absorption.
In the former case, the scaling factor (\ref{eq:6}) is simplified to
$\lambda(x)\approx\frac{x(L-x)}{L\xi}[1-(\frac{x(L-x)}{3\xi_a^2})]$.
For $\lambda(x)\gg 1$ by using the second line of Eq.~(\ref{eq:19}), we find
the local diffusion coefficient
$D(x)\sim e^{-\lambda(x)}$. It enhances from its value in the passive case,
which is $\sim e^{-\frac{x(L-x)}{L\xi}}$ \cite{Tian10},
by a factor of $\sim \exp[\frac{1}{3}(\frac{x(L-x)}{\xi_a\sqrt{L\xi}})^2]$.
The enhancement factor
increases monotonically from the interface to the sample mid-point.
Such inhomogeneous enhancement reflects that
upon turning on absorption resonant states are `killed',
and the portion is determined by the lifetime of resonant states.
Indeed, near the sample mid-point
these states have the longest lifetime
and are most easily to be `killed',
and this accounts for the strongest enhancement at the
mid-point.

The interplay of localization and absorption is even stronger
in the latter case ($\xi_a/L \ll 1$). For
$x$ sufficiently away from the sample center, $x\ll\xi_a/2$,
the scaling factor (\ref{eq:6})
reduces to that of semi-infinite samples,
$\lambda(x)=\frac{\xi_a}{2\xi}(1-e^{-2x/\xi_a})$,
and $D(x)$ is the same as that of semi-infinite samples.
(We consider only $x\leq L/2$ since
$D(x)$ is symmetric with respect to the sample mid-point.)
Its behavior is summarized in Table~\ref{tab:1} and \ref{tab:2}.
Near the sample center, $x\gtrsim \xi_a/2$,
the scaling factor (\ref{eq:6}) is simplified to
$\lambda (x) \approx \lambda (L/2)
$. Thus, $D(x)\approx D(L/2)$ exhibiting a plateau.
Strikingly, despite that the system is controlled by two parameters,
$\xi/L$ and $\xi_a/L$, the plateau value
depends on single parameter, $\xi_a/\xi$. More precisely, Eqs.~(\ref{eq:22})
and (\ref{eq:6}) give Eq.~(\ref{eq:4}).
Moreover, from Eq.~(\ref{eq:19}) we obtain the expression of
$D(L/2)$ for $\xi_a/2\xi\gg 1$ ($\xi_a/2\xi\ll 1$)
which is given by the first line of Table~\ref{tab:1} (\ref{tab:2}).
Note that in the strong absorption limit,
$\xi/\xi_a\rightarrow \infty$, the homogeneity of
the diffusion coefficient is restored, $D(x)=D_0$,
reflecting the absence of localization effects.

That the spatial resolution of $D(x)$ for $x\leq L/2$ is identical to that of
semi-infinite samples when $\xi_a/ L\ll 1$ reflects an important fact:
for strong absorption waves could not propagate from the left-half to the right-half part
of the sample and {\it vice versa}.
As such, resonance states play no roles. Therefore, one might
expect that the phenomenological model of Ref.~\cite{Lagendijk00}
is valid, as observed in numerical experiments \cite{Yamilov13}.

\begin{table}
  \centering
  \caption{Behavior of $D(x)$ at $\xi_a \ll \xi \ll L$}
  \label{tab:2}
\begin{tabular}{|l|l|}
    \hline
    regime & \qquad $D(x)/D(0)$   \\
\hline\hline $x\gtrsim \xi_a/2$ & \qquad $1+c_1 (\xi_a/2\xi)+c_2 (\xi_a/2\xi)^2 +\cdots $ \\
\hline $x\ll \xi_a/2$ & \qquad $1+c_1 (x/\xi)+c_2 (x/\xi)^2+\cdots$ \\
    \hline
  \end{tabular}
\end{table}

In summary, we present the first microscopic theory
showing that in localized open absorbing media photons
display unusual Brownian motion. First of all, the diffusion coefficient is
inhomogeneous in space; most strikingly,
despite that the system is controlled by two parameters ($\xi/L$ and
$\xi_a/L$), it exhibits novel single parameter scaling
namely Eq.~(\ref{eq:22}). The analytic predictions for
the spatial resolution of the diffusion coefficient are
confirmed by numerical simulations. We stress that our theory is
very general and, particularly, valid for arbitrary absorption strength.
In the limiting case of strong absorption
realized experimentally \cite{Yamilov13a},
our results agree well with experimental measurements.
It is very interesting to generalize the present theory to the gained system in the future,
and this may have direct applications in random lasers \cite{Cao02}.

We thank A. Z. Genack and A. G. Yamilov for useful discussions.
This work is supported by the NSFC (No. 11174174) and
by the Tsinghua University Initiative Scientific Research Program (No. 2011Z02151).





\begin{thebibliography}{99}
\bibitem{footnote1} For $\epsilon''\sim {\cal O}(1)$ or in strictly one dimensions
both $D_0$ and $\gamma$ strongly depends on $\epsilon''$. However,
the detailed forms are unimportant for the present work, and therefore
we shall not discuss this issue further.
\end{thebibliography}
\end{document}